\documentclass[preprint,review,12pt]{elsarticle}

\usepackage{amssymb}
\usepackage{graphicx}
\usepackage{threeparttable}

\journal{Journal of Alloys and Compounds}

\begin{document}

\begin{frontmatter}


\title{Anomalous band-gap bowing of AlN$_{1-x}$P$_x$ alloy.}

\author[INT]{M. J. Winiarski}
\author[PWR]{M. Polak}
\author[PWR]{P. Scharoch}
\address[INT]{Institute of Low Temperature and Structure Research, Polish Academy of Sciences, Ok\'olna 2, 50-422 Wroc\l aw, Poland}
\address[PWR]{Institute of Physics, Wroc\l aw University of Technology, Wybrzeze Wyspianskiego 27, 5O-370 Wroc\l aw, Poland}

\begin{abstract}
Electronic structure of zinc blende AlN$_{1-x}$P$_x$ alloy has been calculated from first principles. Structural optimisation has been performed within the framework of LDA and the band-gaps calculated with the modified Becke-Jonson (MBJLDA) method. Two approaches have been examined: the virtual crystal approximation (VCA) and the supercell-based calculations (SC). The composition dependence of the lattice parameter obtained from the SC obeys Vegard's law  whereas the volume optimisation in the VCA leads to an anomalous bowing of the lattice constant. A strong correlation between the band-gaps and the structural parameter in the VCA method has been observed. On the other hand, in the SC method the supercell size and atoms arrangement (clustered vs. uniform) appear to have a great influence on the computed band-gaps. In particular, an anomalously big band-gap bowing has been found in the case of a clustered configuration with relaxed geometry. Based on the performed tests and obtained results some general 
features of MBJLDA are discussed and its performance for similar systems predicted.
\end{abstract}


\end{frontmatter}

\section{Introduction}

Aluminium nitride and phosphide have been drawing an interest due to their present and potential applications in optoelectronics. Particularly attractive is their ability to form mixed systems (alloys) and related opportunity of tuning the band-gap. Zincblende (ZB) phase of AlN exhibits a metastable character, thus experimental studies were performed on the wurtzite type crystal and oriented on using it as a material for ultraviolet solid-state light sources \cite{AlN_Nature}. ZB AlGaP alloys and heterostructures, in turn, are utilized in green-light emission devices \cite{Sonomura1, Ermakov, Sonomura2}. Some technology-related factors, like the problem of miscibility, atomic configuration at a given composition, internal strains, defects etc., and their influence on electronic structure make the issue of forming alloys/heterostructures nontrivial and require careful experimental and theoretical studies. Here, the density functional theory (DFT) based \emph{ab initio} computational methods provide a powerful 
tool since, in principle, they should give an empirical input free  insight into the structural and electronic properties.  Unfortunatelly, the methods have their own limitations, the main of which is the maximum size of the system which can be studied (usually limited to a few tens of atoms). Another known deficiency of DFT/LDA(GGA) (local density and general gradient approximations) is an incorrect value of the band-gap. Thus, there is a continuous need to search for simplifications/approximations which would provide an efficient tool for studying systems' structural and electronic properties, preserving its first principles character.

The electronic and structural properties of ZB AlN and AlP have been extensively investigated \emph{ab initio} \cite{Froyen, Rodriguez, Mujica1, Mujica2, Jivani,Reshak,Annane, Briki, Litimein, Siegel, Karaahanov, Bentouaf}, in the framework of the density functional theory. In this work we focus on the structural and electronic properties of ZB AlN$_{1-x}$P$_x$ alloy. The calculations have been done with the use of the Abinit package \cite{Abinit1, Abinit2}.  The structural optimization has been performed employing the LDA approximation for the exchange-correlation energy functional. The electronic structure calculations have been done by means of the state-of-the-art method, the modified Becke-Jonson potential \cite{MBJLDA} which is known to lead to a satisfactory description of ZB AlN and GaP direct and indirect band-gaps.  The well known super-cell based approach (SC) to calculations has been used as well as the alchemical pseudopotential mixing approximation which is an implementation of virtual crystal 
approximation (VCA) available in the Abinit package. The alchemical mixing approximation is based on the following construction: the local potentials are mixed in the proportion given by mixing coefficients, the form factors of the non-local projectors are all preserved, and all considered to generate the alchemical potential, the scalar coefficients of the non-local projectors are multiplied by the proportion of the corresponding type of atom, the characteristic radius for the core charge is a linear combination of the characteristic radii of the core charges, the core charge function f(r/r$_c$) is a linear combination of the core charge functions. In all the linear combinations the mixing coefficients reflecting the proportion at which particular atoms enter the alloy are used. In the approximation norm conserving pseudopotentials have to be used having the same valence electronic configuration, like e.g. the N and P atoms. A comparison of the results between the supercell (SC) and the alchemical mixing is 
carried out and some specific technical issues are discussed with particular emphasis on the former results for similar systems, e.g. BN$_{1-x}$P$_x$ alloy \cite{BNP}.

\section{Computational details}

As mentioned above the electronic structure calculations for AlN$_{1-x}$P$_x$ alloy have been performed within the density functional theory and the pseudopotential approach. Pseudoatoms have been generated with the use of the Opium package \cite{Opium}. The valence states have been chosen as 2s2p for the N atom and 3s3p for the Al and P atoms. The total energy convergence in the plane wave basis was found to be sufficient for 30 Ha energy cutoff. Firstly, the equilibrium geometries were found via energy/forces relaxation at the Perdew-Wang \cite{LDA} parametrisation of the exchange-correlation energy functional. This was followed by the calculation of the electronic structure employing the MBJLDA method \cite{MBJLDA}. Furthermore, some band-gap results were verified by using a full potential code (FLAPW, Wien2k \cite{Wien2k}). In the SC calculations supercells with 8-atoms (conventional FCC cell) and with 16-atoms ($2\times2\times2$ multiplicity of primitive FCC cell) have been applied, and they led to 
different results of band-gaps. Primitive cell calculations were carried out with $8\times8\times8$ {\bf k}-point mesh with standard shifts for FCC lattice, whereas for supercell systems the  $4\times4\times4$ {\bf k}-point mesh gave sufficient convergence.

\section{Results and discussion}

\begin{figure}
\includegraphics[scale=1.0]{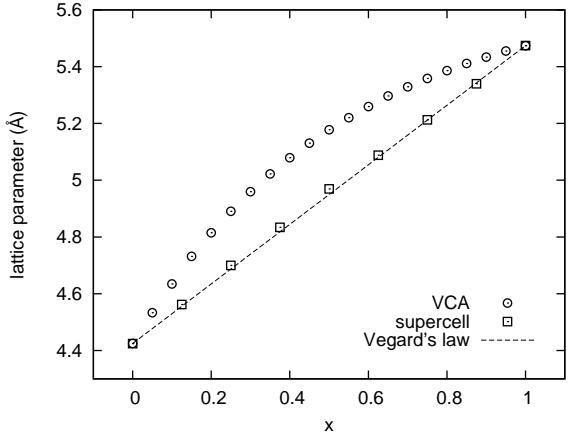}
\caption{Calculated (LDA) lattice parameters of AlN$_{1-x}$P$_x$ alloy.}
\label{Fig1}
\end{figure}

The calculated cell parameters of AlN and AlP, collected in table \ref{table1}, are slightly higher than results reported in the literature. According to our tests this effect is related to the arbitrary selection of the model core (cut-off radius), for the non-linear core-valence exchange-correlation correction. Nevertheless, when compared with the values obtained within the GGA \cite{GGA} calculations, e.g. {\it a} = 5.51 (\AA) for AlP, reported by Annane et al. \cite{Annane}, the cell parameters calculated here  are very satisfactory. The obtained {\it a} = 5.47 (\AA) for AlP is equal to the experimental one \cite{Hellwege} which confirms a good performance of pseudopotentials employed here. However, the composition dependencies of AlN$_{1-x}$P$_x$ alloy cell parameter obtained within VCA and supercell (SC) approaches (depicted in figure \ref{Fig1}), are considerably different. The SC results show linear behaviour (the Vegard's law) while the VCA curve exhibits an anomalous bowing. A similar behavior has 
been reported for BN$_{1-x}$P$_x$ SC calculations \cite{BNP} and needs to be discussed. The correct cell parameter results for alloys are closely related to bonding relaxation between the dopant atoms. In the VCA approach these condition cannot be fulfilled, since, because of the symmetry constraint, no relaxation is allowed within the ZB primitive cell. In the SC calculations a full geometry optimization is performed and it turns out to be crucial for the cell parameters to acquire the correct values (figure \ref{Fig1}). We believe that this is also the reason for the anomalous cell parameter bowing reported for BN$_{1x}$P$_x$ \cite{BNP}.

\begin{figure}
\includegraphics[scale=1.0]{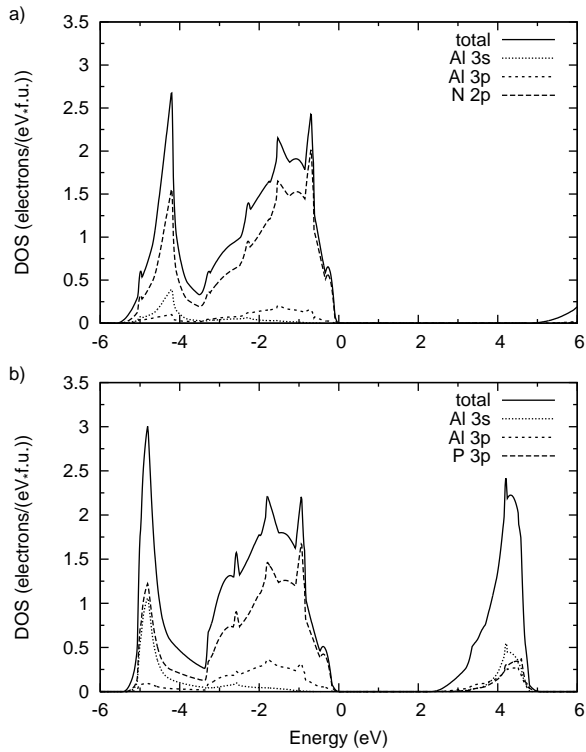}
\caption{Density of states (DOS) plots near the band-gap energy: (a) AlN, (b) AlP.}
\label{Fig2}
\end{figure}

The valence density of states (DOS) for both studied compounds is dominated by N/P 2p/3p states, as illustrated in figure \ref{Fig2}. Aluminium 3p and 3s states contributions in AlN are relatively low and in agreement with former full-potential calculations done for AlN \cite{Litimein}.

\begin{figure}
\includegraphics[scale=1.0]{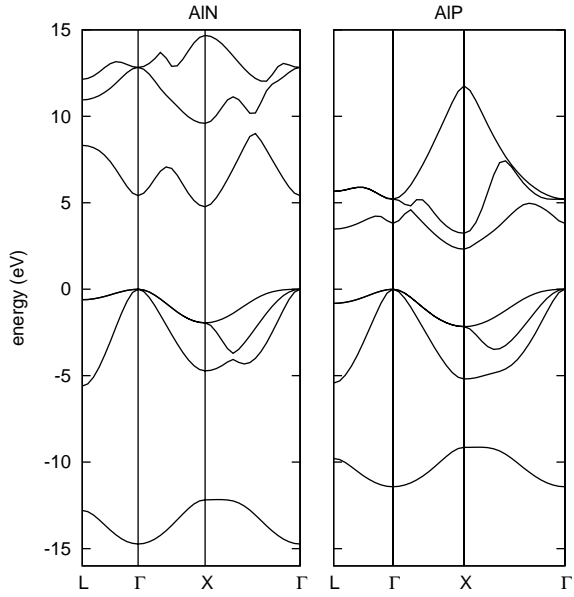}
\caption{Bandplots (MBJLDA) for AlN and AlP.}
\label{Fig3}
\end{figure}

AlN and AlP band structures computed for 8 bands are presented in figure \ref{Fig3}. The overall shape of bands in $L$-$\Gamma$-$X$ direction is the same as in the reported earlier LDA studies \cite{Litimein,Rodriguez} but the MBJLDA band-gaps are significantly bigger. The results of indirect ($\Gamma$-$X$) and direct band-gaps are gathered in table \ref{table2}. The comparison with full-potential data suggests that the semicore states, not included in pseudopotentials (e.g. 2p states for the P atom), do not play a significant role. The agreement between the pseudopotential and the full potential values presented here, and the literature data for the indirect AlP band-gap is excellent. It is also worth noticing, that the outcome of the calculations of the direct AlP band-gap performed on pseudopotentials lacking both semicore states and the non-linear core-valence correlation correction (the latter being an inherent feature of  the MBJLDA method implementation in Abinit code) proved to be closer to the 
experimental data than the full potential calculations. This reveals the character of the MBJLDA method which appears  to work more effectively and accurately with pseudoatoms than with the full potential. In pseudopotentials, a lower than real charge density leads to a more adequate results for the overall band-structure, whereas the fundamental band-gap is almost unchanged. This result should be an inspiration for further careful studies of the MBJLDA approach, in particular in its application to narrow band-gap systems, similar to the study already conducted for compounds with the direct band-gap \cite{Kim}.

\begin{figure}
\includegraphics[scale=1.0]{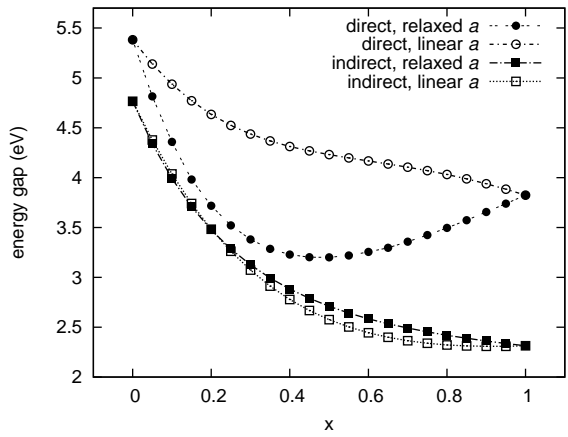}
\caption{Band-gap of AlN$_{1-x}$P$_x$ alloy calculated in the VCA approach.}
\label{Fig4}
\end{figure}

Band-gaps of AlN$_{1-x}$P$_x$ alloy obtained in the VCA method are depicted in figure \ref{Fig4}.  As the band-gap strongly depends on the structural parameters, the anomalous bowing of the relaxed (VCA) lattice parameters (figure \ref{Fig1}) leads to the anomalous results for the direct band-gap, which, at compositions up to $x=0.4$  is close to the indirect one ($\Gamma$-$X$). However, the linear interpolation of lattice parameter between AlN and AlP, which turns out to be in excellent agreement with SC geometry relaxation calculated in this work, enables to obtain an adequate composition dependence of the band-gap. This finding is interesting from the point of view of properties of strained AlN$_{1-x}$P$_x$ heterostructures. Namely, it is evident that the strain induced switching of the fundamental gap  between the indirect and the direct transition is possible, which is crucial for applications of studied here semiconducting material (the enhancement of photoluminescence intensity).

\begin{figure}
\includegraphics[scale=1.0]{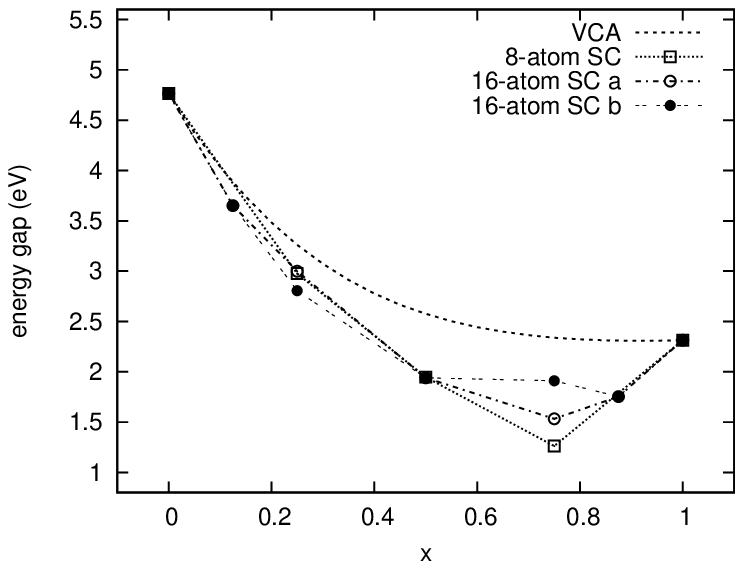}
\caption{Band-gap ($\Gamma$-$X$) of AlN$_{1-x}$P$_x$ alloy calculated with the supercell (8- and 16-atom) approaches without the optimisation of atomic positions: (a) less uniform (quasi-clustered), (b) more uniform P distribution; VCA results from figure \ref{Fig4} added for reference.}
\label{Fig5}
\end{figure}

\begin{figure}
\includegraphics[scale=1.0]{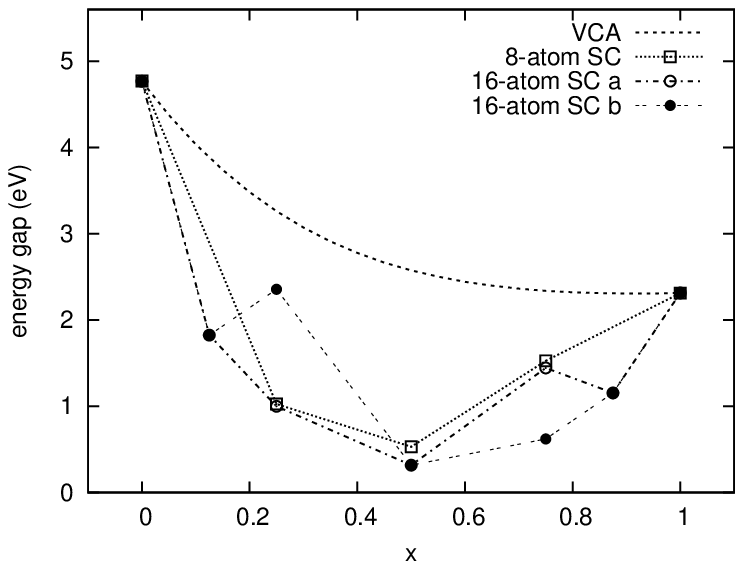}
\caption{Band-gap ($\Gamma$-$X$) of AlN$_{1-x}$P$_x$ alloy calculated with supercell (8- and 16-atom) approach with the optimization of atomic positions: (a) less uniform (quasi-clustered), (b) more uniform P distribution; VCA results from figure \ref{Fig4} added for reference.}
\label{Fig6}
\end{figure}

The SC calculations results of $\Gamma$-$X$ band-gaps for unrelaxed atomic positions are presented in figure \ref{Fig5}. In contrast with the VCA results, the SC band-gap bowings are anomalously high and depend on SC size (8- or 16-atom) and on atoms configuration. The reason for this effect is obvious: the origin of the band-gap lowering is the contribution of particular atomic bonds which are not present (or weakened) in the VCA, because the atomic positions are unrelaxed (the inherent feature of the VCA). The less uniform (quasi-clustered) (a) and the more  uniform (b) P distributions at $x=0.25$ and $x=0.75$ were considered for 16-atom SC. The differences in results between 8- or 16-atom supercell can be attributed to the fact that the smaller is the supercell the less adequate is the simulation of random alloy. This is because in real systems the dopants are distributed randomly even at a high miscibility conditions, the long-range order is lost, and the system is a random alloy. In such situation the 
measured band-gap represents an average over all possible local atomic configurations. In SC simulation at small supercell there exist a supercell-periodic long range order with a particular local atomic configuration.   Moreover, the effect of local configuration is enhanced when the structure is relaxed. This effect is shown in figure \ref{Fig6} and is most distinct at $x$=0.125 and $x$=0.875. Again, such a big band-gap lowering is caused by the contribution of particular bonds which are shrunk or expanded due to the atomic relaxation. A detailed study of this effect is possible by performing the atom/orbital projected calculations of the density of states, and will be a subject of further works. One of methods to correct the description of random alloys  is to use special quasirandom structures (SQS) \cite{Zunger,SQS}. In the case of 8-atom SC the number of possible atomic configurations is very limited, thus the determination of SQS4 \cite{SQS} structure is obvious. However, the result of the band-gap 
bowing for 8-atom SC of AlN$_{0.5}$P$_{0.5}$ is already very high, thus the consideration of further corrections, e.g. using the SQS8 \cite{SQS} seems to be pointless. Furthermore, the band-gap bowing presented here for 8-atom AlN$_{1-x}$P$_x$ (conventional) FCC cell is comparable with MBJLDA results reported for similar system of BN$_{1-x}$P$_x$\cite{BNP}.

The band-gap bowing results (in terms of deviation from Vegard's law) for three compositions of P in AlN$_{1-x}$P$_x$ alloy are collected in table \ref{table3}. Two general conclusions can be drawn: 1) there exist a high asymmetry of the band-gap bowing in both relaxed and unrelaxed structures, and 2) the atomic relaxation leads to large enhancement of the band-gap bowing. Figure \ref{Fig6} reveals rather no systematic behavior of the band-gap under atomic relaxation, apart from the enhacement of bowing. This means that there is a need for the detailed study of this effect.  Similar effects have been observed in BN$_{1-x}$P$_x$ alloy \cite{BNP}, although relatively weaker, however,  in that case  the results might have been also influenced by the non-linear lattice parameter dependence as discussed earlier.  It is also worth considering, that MBJLDA band-gap bowings in alloy systems have not been yet compared with results of other modern methods (e.g. HSE \cite{HSE}), even for systems with less significant 
bowing. This issue requires 
further investigations particularly for systems with such a high band-gap bowing.

\section{Conclusions}

Two approaches: VCA and SC have been explored in this work, together with the MBJLDA approximation in the electronic structure calculations, to study from first principles the composition dependent structural and electronic properties of AlN$_{1-x}$P$_x$ alloy. It has been demonstrated that the VCA poorly reproduces the lattice parameter behavior (deviation from Vegard's law is observed) but, at the imposed \emph{$\acute{a}$ priori} linear dependence of the lattice parameter, it provides a lower limit estimation of the direct and indirect band-gap bowing. On the other hand, the SC calculations of the band-gaps are strongly affected by a particular atomic distribution and their relaxation giving, in some cases, the anomalously large bowings which can be treated as the upper limits of band-gap variations. One of ways of overcoming the difficulties is the special quasirandom structure method.

\section*{Acknoledgements}
The calculations were performed in Wroc\l aw Centre for Networking and Supercomputing.

\begin{table}
\caption{Calculated lattice parameters {\it a} of AlN and AlP, compared with former results.}
\label{table1}
\begin{threeparttable}
\begin{tabular}{lll}
reference &  AlN (\AA) & AlP (\AA) \\ \hline
this work & 4.42 & 5.47 \\
other calc. (LDA) & 4.35$^a$, 4.37$^b$ 4.39$^c$ & 5.41$^d$, 5.44$^a$, 5.45$^e$\\
experimental & - & 5.47$^f$ \\
\end{tabular}
\begin{tablenotes}
\item $^{a-f}$ Ref. \cite{Briki, Litimein, Siegel, Karaahanov, Bentouaf, Hellwege}
\end{tablenotes}
\end{threeparttable}
\end{table}

\begin{table}
\caption{Calculated indirect band-gaps for AlN and AlP (the direct band-gaps in parenthesis), compared with the full potential results; all the values are in eV.}
\label{table2}
\begin{threeparttable}
\begin{tabular}{lll}
reference &  AlN & AlP \\ \hline
pseudopotential & 4.77 (5.38) & 2.31 (3.83)\\
full potential & 4.85 (5.31) & 2.33 (4.30)\\
other calc.$^a$ & - (5.55) & 2.32 (-)\\
experimental $^b$ & - & 2.53 (3.63)\\
\end{tabular}
\begin{tablenotes}
\item $^a$ Ref. \cite{MBJLDA}, $^b$ Ref. \cite{Thompson}
\end{tablenotes}
\end{threeparttable}
\end{table}

\begin{table}
\caption{Calculated band-gap bowing (negative deviation from Vegard's law) for AlN$_{1-x}$P$_x$ ($x=0.25, 0.50, 0.75$); less uniform (quasi-clustered) (a) and more  uniform (b) P distribution; a comparison between VCA and SC results for relaxed and unrelaxed atomic positions; all the values are in eV; for comparison, the direct band-gap bowing for BN$_{1-x}$P$_x$.}
\label{table3}
\begin{threeparttable}
\begin{tabular}{llll}
result / $x$ & $0.25$ & $0.50$ & $0.75$\\ \hline
VCA & 0.89 & 0.95 & 0.59\\
8-atom SC& 1.15 & 1.60 & 1.39\\
relaxed& 3.13 & 3.01 & 1.40\\
16-atom SC (a) & 1.15 & 0.94 & 1.39\\
relaxed& 3.15 & 3.22 & 1.48\\
16-atom SC (b) & 1.35 & 0.94 & 1.02\\
relaxed& 1.80 & 2.69 & 2.30\\
BN$_{1-x}$P$_x$ $^a$& 2.03 & 2.93 & 2.23\\
\end{tabular}
\begin{tablenotes}
\item $^a$ Direct band-gap bowings computed based on data taken from ref. \cite{BNP}.
\end{tablenotes}
\end{threeparttable}
\end{table}

\end{document}